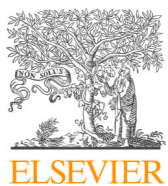
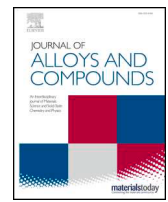

# Effect of small cation occupancy and anomalous Griffiths phase disorder in nonstoichiometric magnetic perovskites

Sagar Ghorai [a],[*], Vitalii Shtender [b], Petter Ström [c], Ridha Skini [a], Peter Svedlindh [a]

[a] *Department of Materials Science and Engineering, Uppsala University, Box 35, SE-751 03 Uppsala, Sweden*
[b] *Department of Chemistry – Ångström Laboratory, Uppsala University, Box 538, Uppsala 75121, Sweden*
[c] *Applied Nuclear Physics, Department of Physics and Astronomy, Uppsala University, Box 516, SE-751 20 Uppsala, Sweden*

A B S T R A C T

The structural, magnetic, magnetocaloric and Griffiths phase (GP) disorder of non-stoichiometric perovskite manganites $La_{0.8-x}Sr_{0.2-y}Mn_{1+x+y}O_3$ are reported here. Determination of valence states and structural phases evidenced that the smaller cations $Mn^{2+}$ and $Mn^{3+}$ will not occupy the A-site of a perovskite under atmospheric synthesis conditions. The same analysis also supports that the vacancy in the A-site of a perovskite induces a similar vacancy in the B-site. The $La^{3+}$ and $Sr^{2+}$ cation substitutions in the A-site with vacancy influences the magnetic phase transition temperature ($T_C$) inversely, which is explained in terms of the electronic bandwidth change. An anomalous non-linear change of the GP has been observed in the Sr-substituted compounds. The agglomeration of $Mn^{3+}$-$Mn^{4+}$ pairs (denoted as dimerons), into small ferromagnetic clusters, has been identified as the reason for the occurrence of the GP. A threshold limit of the dimeron formation explains the observed non-linear behaviour of the GP formation. The Sr-substituted compounds show a relatively large value of isothermal entropy change (maximum 3.27 J/kgK at $\mu_0 H = 2T$) owing to its sharp magnetic transition, while the broad change of magnetization in the La-substituted compound enhances the relative cooling power (maximum 98 J/kg at $\mu_0 H = 2T$).

## 1. Introduction

The perovskite manganites $ABO_3$ (where the B-site is occupied by Mn) are considered as one of the most strongly correlated electron systems exhibiting a rich phase behaviour due to strong interactions between charge, spin, orbital and lattice degrees of freedom. Substitution in the A-site by smaller cations ($Mn^{2+}$, $Sc^{3+}$, $In^{3+}$, etc.) often results in a large lattice distortion and distinct structural and magnetic properties [1]. Applying a high-pressure synthesis method it has been proved that small cations like $Mn^{2+}$ and $Mn^{3+}$ can occupy the A-site of a perovskite [1–3]. Moreover, there have been many reports [4–18] where the analysed material synthesized at **n**o **e**xternal **p**ressure (**nep**), with a general formula $A_{1-\Delta}Mn_{1+\Delta}O_3$ (e.g. $(Nd_{0.7}^{3+}Sr_{0.3}^{2+})_{1-x}Mn_{1+x}O_3$ [13]), exhibits an excess of Mn and the same amount of vacancy in the A-site. The question naturally arises: without any extra energy (e.g. due to external pressure [1–3]) will the smaller Mn cation occupy the A-site or will it make a separate phase? Suspiciously, most of the **nep** publications (some did not mention or present results from X-ray diffraction or magnetization measurements) report a secondary $Mn_3O_4$-phase, which can be a result of over-stoichiometric Mn ions not occupying the A-site. Moreover, it is known that the $Mn_3O_4$-phase can form due to synthesis above 1273 K [19]. To avoid the confusion relating to synthesis conditions, in this work a parent perovskite with less than 1% of $Mn_3O_4$-phase and its A-site substituted compound ($A_{1-\Delta}Mn_{1+\Delta}O_3$) have been synthesized under identical conditions. From the comparative analysis of structural phase formation, the A-site occupation probability for the smaller cation $Mn^{2+}/Mn^{3+}$ has been determined.

In manganite perovskites, the antiferromagnetic super-exchange interaction ($Mn^{3+}$-$O^{2-}$-$Mn^{3+}$ or $Mn^{4+}$-$O^{2-}$-$Mn^{4+}$) and ferromagnetic double exchange interaction ($Mn^{3+}$-$O^{2-}$-$Mn^{4+}$) depend upon the availability of Mn-ions with different oxidation states [20]. As discussed before, owing to the absence of external pressure, if the smaller cations $Mn^{2+}/Mn^{3+}$ do not occupy the A-site, a simple vacancy will be created in the A-site of the compound and the A-site as well as the B-site ionic ratio will be affected. Thus, the B-site ionic ratio will vary depending upon the occupancy of smaller cations ($Mn^{2+}/Mn^{3+}$) or vacancies in the A-site. In our previous work [20], we have discussed the effect of Jahn-Teller (JT) active ions ($Mn^{3+}$ and

[*] Corresponding author.
*E-mail address:* sagar.ghorai@angstrom.uu.se (S. Ghorai).

$Cu^{2+}$) on the magnetic disorder (Griffiths phase formation). In the current work, the A-site occupancy (with Mn-ions or vacancies) will influence the number of JT-active ions and hence the magnetic disorder. Using X-ray photoelectron spectroscopy, the valence states of Mn-ions in the compounds were determined and the effect on magnetic disorder phase formation has been investigated.

Apart from the question of the Mn-occupancy and extra phase formation, the A-site of perovskites exhibits mixed valence states, $A'^{3+}_{1-x}A''^{2+}_{x}MnO_3$, where $A'$ is a cation with +3 oxidation state (e.g. $La^{3+}$, $Pr^{3+}$, etc.) and $A''$ is a cation with +2 oxidation state (e.g. $Ca^{2+}$, $Sr^{2+}$, $Pb^{2+}$, etc.). With this mixed valance condition, if an extra $\Delta$ amount of Mn is added, then the same amount of vacancy in the A-site can be created in two ways, either forming $A'^{3+}_{1-x-\Delta}A''^{2+}_{x}Mn_{1+\Delta}O_3$ or $A'^{3+}_{1-x}A''^{2+}_{x-\Delta}Mn_{1+\Delta}O_3$ types of compounds. Noticeable, for both types the ratio of total amounts of A-site and B-site cations ($\frac{1-\Delta}{1+\Delta}$) is the same. However, as there is a size mismatch between the +3 and +2 cations it will influence the overall ionic bond lengths and angles between the magnetic ions in the compound, hence influencing the magnetic interactions and the magnetic state of the compound. Recently, in the $La_{0.8-x}K_xMn_{1+x}O_3$ system, a slight increase of the magnetic phase transition temperature ($T_C$) was observed with +3 cation substitution [21]. It was argued that the change of the electronic band width (will be discussed later) is responsible for this change of $T_C$. A similar variation of the electronic band width has been observed for the different compounds studied in this work. Moreover, in this work for the first time (to the best of our knowledge) it has been observed that +3 and +2 cation substitution with vacancy formation in the A-site tunes the magnetic phase transition temperature ($T_C$) in opposite ways, i.e. with +3 cation substitution $T_C$ increases, while it decreases with +2 cation substitution comparing with the parent compound. Apart from the tuning of $T_C$, this work describes a unique way of tuning the isothermal entropy change without insertion of any new element in the compound, which will be important for near room temperature magnetic refrigeration applications.

## 2. Material design and experimental details

In this work, we have prepared phase pure $La_{0.8}Sr_{0.2}MnO_3$ as a parent compound. In order to investigate the effect of excess Mn and the same amount of vacancy formation in the A-site, we have prepared a series of compounds with the general formula $La_{0.8-x}Sr_{0.2-y}Mn_{1+x+y}O_3$, where $x$ and $y$ separately take the values of 0, 0.1 and 0.15. To summarize, the following compounds have been investigated in this work: $La_{0.8}Sr_{0.2}MnO_3(x=0, y=0)$, P; $La_{0.7}Sr_{0.2}Mn_{1.1}O_3(x=0.1, y=0)$, L10; $La_{0.65}Sr_{0.2}Mn_{1.15}O_3(x=0.15, y=0)$, L15; $La_{0.8}Sr_{0.1}Mn_{1.1}O_3(x=0, y=0.1)$, S10; and $La_{0.8}Sr_{0.05}Mn_{1.15}O_3(x=0, y=0.15)$, S15. The abbreviated names P, L10, L15, S10 and S15 of the above-mentioned compounds will be used through-out this paper.

The five compounds were synthesized by a sol-gel process [22]. Calculated amounts of high purity $La(NO_3)_2 \cdot 6H_2O$, $Sr(NO_3)_2 \cdot H_2O$ and $C_4H_6MnO_4 \cdot 4H_2O$ were dissolved in di-ionized water as the starting material. The later synthesis steps were as described by Manuel et al. [22]. The final sintering temperature was 1373 K followed by several intermediate grinding and heating steps. The X-ray powder diffraction (XRPD) data was collected by using a Bruker D8 Advance diffractometer with Cu-K$_\alpha$ radiation at room temperature. XRPD diffractograms were collected in a wide range (17–140°) and with an angle step size of 0.02°. XRPD data were analyzed using the Rietveld method implemented in the Topas 6 Academic software [23,24]. The chemical properties of the compounds were characterized by Rutherford backscattering spectrometry (RBS) with a 2 MeV $^4He^+$ beam, and a solid state silicon detector with 15 keV energy resolution placed at a backscattering angle of 170°. For quantification of the Mn-oxidation states, X-ray photoelectron spectroscopy (XPS) data was collected by using a "PHI Quantera II" system with an Al-K$\alpha$ X-ray source and a hemispherical electron energy analyser having a pass energy of 26.00 eV. Before recording an XPS spectrum the sample surface was cleaned by 200 eV Ar-ion sputtering for 30 s. All magnetic measurements were carried out with a Quantum Design MPMS XL, with a highest applied magnetic field of 5 T.

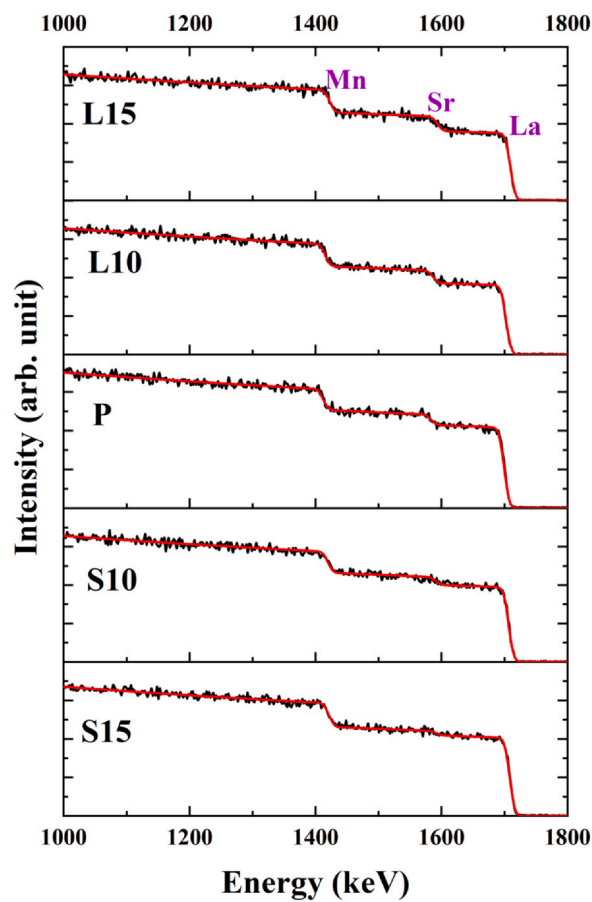

**Fig. 1.** $^4$He RBS data with SIMNRA fitted spectra (red solid lines) for the studied compounds.

## 3. Results and discussions

### 3.1. Chemical composition analysis

In order to verify the presence of extra Mn in the studied compounds and the desired chemical stoichiometry, ion beam analysis has been performed. As discussed previously, the magnetic properties of the compounds will depend on the occupancy of the extra Mn. The $^4He^+$ RBS data of the different compounds are shown in Fig. 1. The experimental data were analysed using the SIMNRA [25] software by adapting the concentrations of Mn, Sr and La while assuming 60 at% of O, and reproducing the experimental signal heights in regions of interest associated with each adapted element to +−0.5%. These regions were selected such that the results are representative of the samples' composition between depths of 50 nm and 100 nm beneath the surface. Careful inspection of the graphs in Fig. 1, gives an indication of increasing La concentration with depth within the first 100 nm, and as such the values given in Table 1 may slightly underestimate the bulk La fraction. A fraction of Mn elevated above 20 at% is indicated in all samples where such an increase was intended. The increase is, however, marginally smaller than expected, especially in the La-deficient sample L15 where the Sr fraction is also elevated in the probed depth range.

**Table 1**
Atomic fractions of the cations in the compounds, calculated from RBS data.

|  |  | Samples | | | | |
|---|---|---|---|---|---|---|
|  |  | S15 | S10 | P | L10 | L15 |
| **La (at%)** | Observed | 15.2(1) | 15.2(1) | 15.6(1) | 13.3(1) | 12.8(1) |
|  | Expected | 16 | 16 | 16 | 14 | 13 |
| **Sr (at%)** | Observed | 1.8(1) | 3.5(2) | 4.4(2) | 5.3(2) | 6.1(2) |
|  | Expected | 1 | 2 | 4 | 4 | 4 |
| **Mn (at%)** | Observed | 23.0(2) | 21.3(2) | 20.0(2) | 21.4(2) | 21.1(2) |
|  | Expected | 23 | 22 | 20 | 22 | 23 |

### 3.2. Structural properties and Mn-site preference

The crystallographic phase analysis was performed by Rietveld refinement[26] of the XRPD spectra collected at room temperature (see supplementary information). The parent compound (P) shows a pure rhombohedral phase along with a negligible amount (< 1 at%) of tetragonal $Mn_3O_4$ phase (see Fig. 2(a)). In the Sr-substituted compounds (S10 and S15), apart from the main rhombohedral phase, a significant amount of $Mn_3O_4$ phase has been observed, which is proportional to the degree of Sr-substitution. Similar proportional increment of the $Mn_3O_4$ phase was also observed in the La-substituted compounds (L10 and L15). It is to be noted that the atomic substitution in the La-site creates an additional orthorhombic phase (cf. Fig. 2(a)).

From the comparison of La-site and Sr-site substitutions, the $La^{3+}$ cation substitution leads to co-existence of the rhombohedral and orthorhombic phases. Radaelli et al. [27] has reported a phase transition from orthorhombic to rhombohedral phase with increasing average A-site atomic radius $\langle r_A \rangle$. To calculate $\langle r_A \rangle$ for our compounds, it is necessary to identify which atoms apart from La and Sr are present in the A-site. The calculated values of $\langle r_A \rangle$, with three possible occupancies of $Mn^{2+}$ and $Mn^{3+}$ (smaller cations) in the A-site are depicted in Fig. 2(b). If there is no Mn-ion in the A-site or if only $Mn^{2+}$ is present in the A-site, the calculated values of $\langle r_A \rangle$ are exactly same for the La-substituted compounds. Similarly, for Sr-substituted compounds the same values of $\langle r_A \rangle$ are observed for the case of no vacancy in the A-site and only $Mn^{2+}$ in the A-site. Moreover, for every possible A-site occupancy, the value of $\langle r_A \rangle$ decreases with La-site or Sr-site substitutions, with respect to the parent compound. The decrease of $\langle r_A \rangle$ can be related to the increase of orthorhombic phase in the La-substituted compounds, as expected from the work of Radaelli et al. [27].

As discussed before, there is some confusion as to the substitution of larger A-site cations with smaller Mn-ions. In this situation, the previously mentioned three possible occupation states for the extra Mn-ions in the compounds are described in the following models (only ionic bonds are considered in these models):

**Model 1 (M1).** The amount of +3 and +2 vacancy in the A-site will be compensated by $Mn^{3+}$ and $Mn^{2+}$ ions. In this way there will be no extra Mn-phase formed and the result can be expressed as,

$La_{0.8-x}Sr_{0.2-y}Mn_{1+x+y}O_3 \rightarrow (La^{3+}_{0.8-x}Mn^{3+}_x Sr^{2+}_{0.2-y}Mn^{2+}_y)(Mn^{3+}_{0.8}Mn^{4+}_{0.2})O_3.$

The valance state of Mn is calculated as,

$$v = (3.2 + 3x + 2y)/(1 + x + y). \quad (1)$$

**Model 2.** In this model we consider that there will be no Mn-ions occupying the A-site. As a result, the extra Mn will form the $Mn_3O_4$ phase. In this scenario, two cases can arise.

**Case 1. of model 2 (M2C1):** With no Mn-ions occupying the A-site there will be a vacancy created in the A-site, while there will be no vacancy in the B-site. Thus, the system will be a perovskite structure with vacant A-site along with an extra Mn-phase, which can be expressed as,

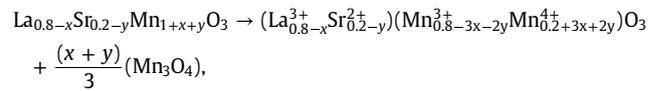

implying that the amount of $Mn_3O_4$ phase becomes,

$$Mn_3O_4 \text{ phase \%} = \frac{1}{1 + \frac{3}{x+y}} \times 100\%. \quad (2)$$

Similar to Eq. 1, the Mn valence state becomes $v = (3.2 + \frac{17}{3}x + \frac{14}{3}y)/(1 + x + y)$.

**Case 2. of model 2 (M2C2):** In this case, along with the vacancy in the A-site, the same amount of vacancy is considered in the B-site. Thus, the A- and B-sites will have same number of atoms. In terms of chemical reaction, the Mn-ions will make ionic bonds with the available A-site cations to form a perovskite phase and the rest of the Mn will form the $Mn_3O_4$ phase. This can be expressed as, and the Mn valence state becomes $v = (3.2 + \frac{7}{3}x + \frac{4}{3}y)/(1 + x + y)$. Similar to Eq. 2, $Mn_3O_4$ phase % = $\frac{1}{1 + \frac{3}{2(x+y)}} \times 100\%$.

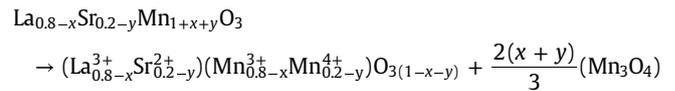

**Model 3 (M3).** According to this model, only $Mn^{2+}$ will occupy the A-site owing to the smaller difference in atomic radius between $Sr^{2+}$ (1.31 Å) and $Mn^{2+}$ (0.83 Å) as compared to the difference between $La^{3+}$ (1.216 Å) and $Mn^{3+}$ (0.645 Å). The rest of the Mn will form the $Mn_3O_4$ phase. The $Mn^{2+}$ A-site occupation has also been observed for La-deficient $LaMnO_3$ [28]. Similar to Model 2, also here are two cases considered.

**Case 1. of model 3 (M3C1):** In contrast to **M2C1**, in this model the A-site Sr-vacancy is compensated by $Mn^{2+}$, which can be expressed as, with $v = (3.2 + \frac{17}{3}x + 2y)/(1+x+y)$ and $Mn_3O_4$ phase % = $\frac{1}{1+\frac{3}{x}} \times 100\%$.

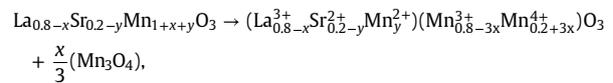

**Case 2. of model 3 (M3C2):** In this model the perovskite phase has the same amount of A- and B-site atoms along with $Mn^{2+}$ compensating for the Sr-vacancy in the A-site, which can be expressed as, with $v = \left(3.2 + \frac{7}{3}x + 2y\right)/(1+x+y)$ and $Mn_3O_4$ phase % = $\frac{1}{1+\frac{3}{2x}} \times 100\%$.

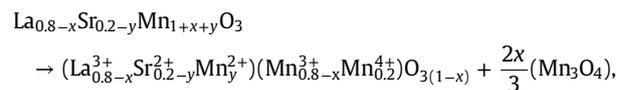

From this, it is clear that for every model system, the valence state of Mn and the amount of $Mn_3O_4$ phase are unique. Thus, to identify the proper model system, the above-mentioned calculated values are compared with the corresponding experimentally determined values. The valence states of Mn for the different compounds have been determined using the XPS spectra (see supplementary). As discussed in our previous work [20], due to the difficulty of Mn 2p peak fitting for $La_{1-x}Sr_xMnO_3$ compounds, only the Mn 3s states have been considered. Owing to the parallel and antiparallel coupling of the Mn3s core-hole and the Mn 3d electron spins, two distinct peaks appear in the XPS spectra of the Mn 3s state (see supplementary information). The binding energy splitting ($\Delta E$) between these two peaks and the valence state ($v$) of Mn follows a linear relationship [29],

$v = A - B\cdot\Delta E.$

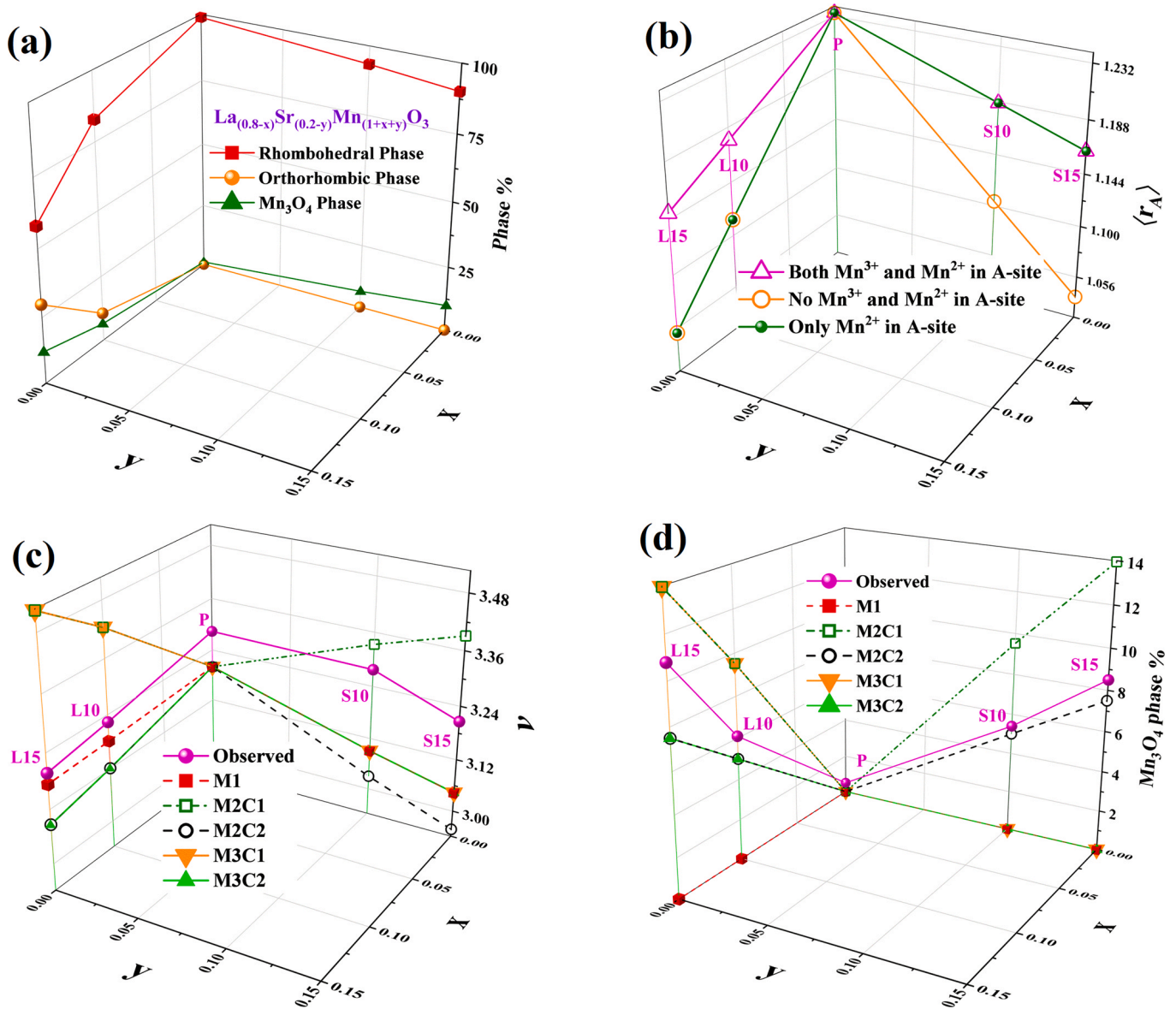

**Fig. 2.** (a) Phase abundance from XRPD analysis. (b) Calculated values of ⟨$r_A$⟩ for possible occupations of Mn-ions in the A-site. (c) Experimentally observed and model calculated Mn valence states and (d) experimentally observed and model calculated amount of $Mn_3O_4$ phase; see main text for model descriptions.

Using values of the reference samples $Mn_2O_3$ and MnO, the values of A and B could be determined as 11.04 and 1.45 eV$^{-1}$, respectively. The calculated $v$ for the different compounds in different model systems along with the observed values are shown in Fig. 2(c). The decreasing trend of the valence states with respect to the parent compound is obvious for the M1, M2C2 and M3C2 model systems. This indicates that either the smaller cations ($Mn^{2+}$ and $Mn^{3+}$) have completely occupied (model M1) the A-site, or the vacancy created in the A-site creates a similar vacancy in the B-site (models M2C2 and M3C2).

For the further confirmation of the models, the amount of $Mn_3O_4$ phase formation have been calculated as described in Eq. (2) for every model system. The calculated values of $Mn_3O_4$-phase from the above-mentioned models and the observed $Mn_3O_4$-phase from XRPD analysis are plotted in Fig. 2(d). From the figure, it is clear that the M2C2 model, which is the perovskite structure with same amount of A- and B-site atoms along with no Mn-atoms occupying the A-site, matches well with the experimental values.

It should be kept in mind that owing to the low intensity of the Mn 3s peaks the deconvolution of the peaks contains an error of maximum 9% and the phase analysis from XRPD contains an error of maximum 2%. Thus, the above analysis should be considered as a qualitative analysis yielding information about the Mn-occupancy. However, from this comparative analysis, it is quite clear that smaller cations will not occupy the A-site for samples prepared under **nep** conditions.

### 3.3. Magnetic phase transition and magnetocaloric properties

The temperature dependence of the low field magnetization is shown in Fig. 3(a) for the different compounds. The paramagnetic to ferromagnetic transition temperature $T_C$ shows opposite trends for

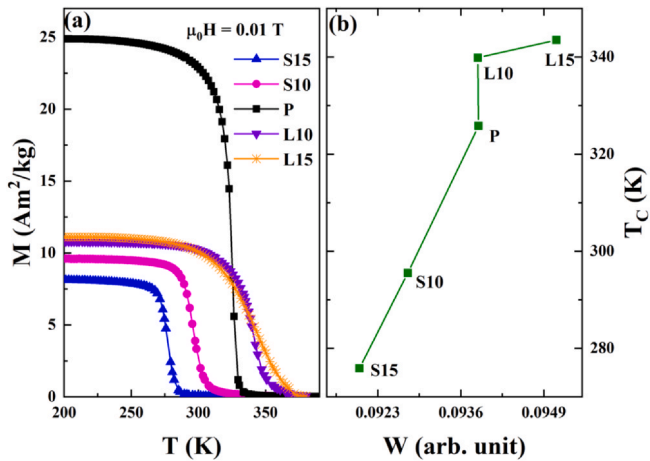

**Fig. 3.** (a) Temperature dependent magnetization and (b) $T_C$ as a function of charge carrier bandwidth.

**Table 2**
Comparative values of isothermal entropy change and relative cooling power for different manganite oxides with transition temperature near room temperature.

| Sample | $T_C$ (K) | $\mu_0 H_{max}$ (T) | $-\Delta S_M^{max}$ (J/kgK) | RCP (J/kg) | References |
|---|---|---|---|---|---|
| S15 ($La_{0.8}Sr_{0.05}Mn_{1.15}O_3$) | 275.9 | 2 | 3.27 | 81.71 | This work |
| S10 ($La_{0.8}Sr_{0.1}Mn_{1.1}O_3$) | 295.5 | 2 | 2.68 | 89.66 | This work |
| P ($La_{0.8}Sr_{0.2}MnO_3$) | 325.8 | 2 | 2.92 | 86.32 | This work |
| L10 ($La_{0.7}Sr_{0.2}Mn_{1.1}O_3$) | 339.8 | 2 | 1.93 | 97.68 | This work |
| L15 ($La_{0.65}Sr_{0.2}Mn_{1.15}O_3$) | 343.5 | 2 | 1.31 | 93.82 | This work |
| $La_{0.8}K_{0.2}MnO_3$ | 330 | 1 | 1.72 | 34 | [21] |
| $La_{0.75}K_{0.2}Mn_{1.05}O_3$ | 332 | 1 | 1.79 | 32 | [21] |
| $La_{0.7}K_{0.2}Mn_{1.1}O_3$ | 332 | 1 | 1.52 | 34 | [21] |
| $La_{0.8}K_{0.1}MnO_3$ | 300 | 2 | 1.65 | 95.81 | [38] |
| $La_{0.75}Ca_{0.05}Na_{0.2}MnO_3$ | 300 | 2 | 3.12 | 90 | [39] |
| $La_{0.5}Pr_{0.2}Ca_{0.1}Sr_{0.3}MnO_3$ | 296 | 2 | 1.82 | 146.5 | [19] |
| $La_{0.4}Pr_{0.3}Ca_{0.1}Sr_{0.3}MnO_3$ | 289 | 2 | 3.08 | 83.3 | [40] |
| $La_{0.8}Ca_{0.2}MnO_3$ | 236 | 2 | 5.96 | 112.36 | [41] |
| $La_{0.8}Na_{0.1}\square_{0.1}MnO_3$ | 295 | 2 | 2.97 | 96.06 | [42] |
| Gd | 299 | 2 | 4.20 | 196 | [43] |

La-site and Sr-site substitutions. With increasing La-site substitution (in the L10 and L15 compounds) $T_C$ increases while it decreases with increasing Sr-site substitution (in the S10 and S15 compounds) as compared to the parent compound (P). Noticeable, Sr-site substitution results in an almost linear decrease of $T_C$ (see Table 2) with increasing substitution, in contrast to the non-linear increase of $T_C$ with La-site substitution. Moreover, the magnetic phase transition broadens with La-substitution. The decrease or increase of $T_C$ indicates a corresponding decrease or increase of the overall strength of the magnetic interactions between different magnetic ions. In the perovskite manganites, the magnetic interaction strength can be related to the metal (3$d$ orbitals) and O (2$p$ orbitals) orbital overlap, influenced by the bond length and bond angle between neighbouring Mn atoms in the compound. Moreover, the charge carrier bandwidth $W$ can be expressed in terms of these structural parameters as [30],

$$W \propto \frac{\cos(1/2(\pi - \langle Mn - O - Mn \rangle))}{d_{Mn-O}^{3.5}} \quad (3)$$

where $\langle Mn - O - Mn \rangle$ and $d_{Mn-O}$ are the bond angle and bond length, respectively.

A proportionality between $T_C$ and $W$ has frequently been observed in perovskite manganites [31,32]. The calculation of $W$ in this work is not straight forward owing to the presence of different bond lengths and bond angles corresponding to different structural phases, especially for the L10 and L15 samples.

Fig. 3(b) shows $T_C$ versus $W$ calculated from Eq. (3). In this calculation the bond angles and bond lengths for the different structural phases were used weighted with their respective phase percentage. A linear variation of $T_C$ is observed for the Sr-substituted compounds, which is in a good agreement with previously reported results on different manganite perovskites [31,32]. The dependence for the La-substituted compounds is more complex, which could be due to the complexity arising in the calculation of $W$ in the presence of two major phases.

The isothermal entropy change ($\Delta S_M$) can be used to investigate the potential of a magnetic compound for use in magnetocaloric effect (MCE) cooling devices. Using one of Maxwell's thermodynamic relations, $\Delta S_M$ for a change of magnetic field from 0 to $\mu_0 H_{max}$ at constant temperature $T$ is calculated as [33,34],

$$\Delta S_M(T, \mu_0 H) = \int_0^{\mu_0 H_{max}} \left(\frac{\partial M}{\partial T}\right)_{\mu_0 H} d(\mu_0 H).$$

The temperature dependent isothermal entropy change for the five compounds are shown in Fig. 4(a) and Table 2. From the observed temperature dependence of $\Delta S_M$, it is clear that apart from a decrease of $T_C$ with Sr-site substitution there is a minimal change of $\Delta S_M$ amplitude while the amplitude decreases significantly with La-site substitution. This can be explained from the sharpness of the magnetic transition (cf. Fig. 3(a)) as well as from the full width at half maximum of $\Delta S_M(T)$ (cf. Fig. 4(a)). A sharp magnetic transition implies a large change of the isothermal magnetization for a given applied field, which explains the high value of isothermal entropy change [35] observed for the S15 compound. It is known that first order magnetic transitions exhibit sharp transitions and therefore large $\Delta S_M$. Thus, to identify the order of the magnetic transition, the Banerjee criterion, based on Arrott plots ($M^2$ versus $H/M$) has been analysed (not shown here) [36–38]. The absence of negative slopes in the Arrott plots confirms the transitions to be of second ordered for all five compounds. However, the Banerjee criterion is based on a mean field approximation, which may not be valid for a real compound. Recently, J. Y. Law et al. [37] suggested a quantitative fingerprint for a first order magnetic transition based on the exponent $n'$ defined as [36,37],

$$n'(T, \mu_0 H_f) = \frac{dln(|\Delta S_M(T, \mu_0 H_f)|)}{dln(H_f)}.$$

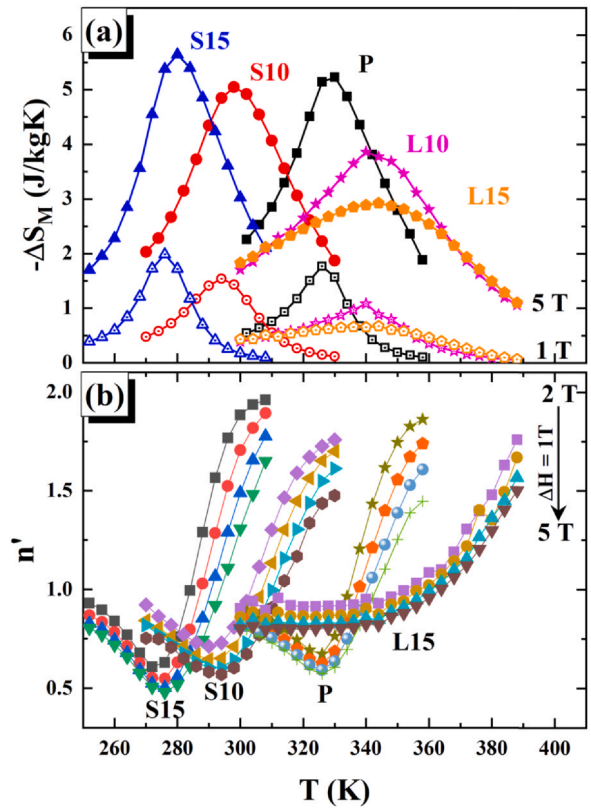

**Fig. 4.** (a) Isothermal entropy change for the different compounds versus temperature for 1 T and 5 T magnetic field changes. (b) Temperature and magnetic field dependence of the exponent n' for the different compounds. L10 and L15 results overlap and the L10 curve is therefore not shown here.

A value of $n' > 2$ near $T_C$ is a fingerprint of a first order magnetic transition while for both types of transitions, the $n'$ value tends to 1 at $T \ll T_C$ and approaches 2 at $T \gg T_C$. In Fig. 4(b), the values of $n'$ are shown for all compounds (except L10 for better visualization, as results of L10 and L15 overlap each other). In the analysis, small fields are excluded in order to avoid a multi domain effect [37]. The absence of $n' > 2$ near $T_C$ confirms a second order magnetic transition for all compounds.

Another important parameter for a material to be used as refrigerant in a magnetocaloric cooling device is the relative cooling power (RCP), defined as [19],

$$RCP = -\Delta S_M^{max} \times \Delta T_{FWHM},$$

where $-\Delta S_M^{max}$ is the maximum value of the isothermal entropy change and $\Delta T_{FWHM}$ is the full width at half maximum of $\Delta S_M(T)$. The observed values of RCP along with the values of the isothermal entropy change are listed in Table 2. With comparatively large values of RCP and $-\Delta S_M^{max}$, the La and Sr-substituted compounds describe an important class of MCE materials near the room temperature.

### 3.4. Induced Griffiths phase disorder and Jahn-Teller active ions

A deviation from the Curie-Weiss law provides evidence of magnetically ordered spin-clusters in the paramagnetic region. This magnetic inhomogeneity phase known as the Griffiths phase (GP), persists up to a certain temperature $T_G$ (Griffiths temperature) above $T_C$ and the region between $T_C$ and $T_G$ is known as the Griffiths phase region [44]. In perovskite manganite systems, the existence of the GP is often connected to the presence of Jahn-Teller (JT) active $Mn^{3+}$ ions [45]. In our previous work [20], we have shown that the presence of JT-active ions influences the dimeron ($Mn^{3+}$-$Mn^{4+}$ pair) formation, which gives rise to the GP singularity. A dimeron is a pair of $Mn^{3+}$-$Mn^{4+}$ ions, which has reduced JT-distortion compared to an isolated $Mn^{3+}$ ion and agglomeration of dimerons in a lattice gives rise to small ferromagnetic clusters, which results in the GP formation. In this work, all compounds except the parent compound (P) show (Fig. 4(**a**)) GP behaviour. For a better comprehension of the GP the temperature dependence of the magnetization has been fitted with the characteristic temperature dependent GP equation [45],

$$\chi^{-1} \propto (T - T_C^R)^{1-\lambda}, \quad (4)$$

where $0 \leq \lambda < 1$ and $T_C^R$ is the critical temperature where the susceptibility tends to diverge. The GP-region can be quantified as,

$$GP\% = \frac{T_G - T_C}{T_C} \times 100. \quad (5)$$

**Table 3**
Fitted parameters to the Griffiths phase equation for different compounds.

| Sample | $T_C$ (K) | $T_G$ (K) | GP% | $\lambda$ |
|---|---|---|---|---|
| S15 | 275.9 | 308.5 | 11.8 | 0.24(2) |
| S10 | 295.5 | 377.4 | 27.7 | 0.79(1) |
| P | 325.8 | – | 0 | 0 |
| L10 | 339.8 | 384.1 | 13.0 | 0.47(2) |
| L15 | 343.5 | 382.9 | 11.5 | 0.45(1) |

The extracted values of GP% and $\lambda$ are listed in Table 3 and shown in Fig. 5 (b)-(c). From the extracted values it is obvious that, in the La-substituted compounds the GP-region has increased compared to the parent compound, however it does not vary much with the amount of La-substitution. On the other hand, a strong dependency of the GP-region on the amount of substitution has been observed in the Sr-substituted compounds. Interestingly, the GP-region increased initially with Sr-substitution, however further increase of the substitution, results in a decrement of the GP-region. To understand this behaviour the ratio of JT active $Mn^{3+}$ to non-JT active $Mn^{4+}$ ions was determined for the perovskite phase using model M2C2 and shown in Fig. 5(d). In this ratio calculation the $Mn^{3+}$ ions present in the $Mn_3O_4$ phase are not considered as they create antiferromagnetic bonds with the $Mn^{2+}$-ions and hence will not be available for ferromagnetic cluster formation. From Fig. 5(d), the amount of $Mn^{3+}$ cations increases with increasing Sr-site substitution, while it varies inversely with La-site substitution. The decrease of $Mn^{3+}$ cations in the La-substituted compounds is very small, which can explain the similar size of the GP-region in the L10 and L15 compounds.

To explain the relationship between GP formation and the ratio of $Mn^{3+}$ to $Mn^{4+}$ ions, we have considered a model system (cf. Fig. 5(e) to (g)) following the model suggested by L. Downward et al. [46]. In Fig. 5(e), a rhombohedral (as this phase is the primary phase for the studied compounds, however any other system will also satisfy this model) unit cell has been considered for the $La_{0.8}Sr_{0.2}MnO_3$ system (parent compound), where only the B-site Mn-atoms are shown in the $ac$ plane. The $Mn^{3+}$ ($Mn^{4+}$) ions describe the JT distorted (non-JT distorted) lattice sites. Thus, the model system in Fig. 5(e), represents the compound P, without any GP. First, we will consider the Sr-substituted compounds. Sr-substitution in the parent compound implies replacement of $Mn^{4+}$ ions with $Mn^{3+}$ ions (see the ionic ratio change in Fig. 5(d)). To see the effect of this replacement, three $Mn^{4+}$ ions are indicated in Fig. 5(e) and the replacement with $Mn^{3+}$ ions are shown in Fig. 5 (f) and (g). The replacement of the $Mn^{4+}$ ions results in the formation of small dimerons (shown with blue circles) and the agglomerated dimerons form ferromagnetic clusters (shown with green curve). The transition from Fig. 5(f) to (g) indicates that, if the number of $Mn^{3+}$ ions cross a threshold limit (where the number of dimerons is maximum), the ferromagnetic clusters will diminish in size and hence the GP-region will be reduced. This strongly supports the observed GP behaviour in the Sr-substituted compounds. The GP initially increases in the S10 compound, further increase of $Mn^{3+}$ ions in the S15 compound crosses the above mentioned dimeron-threshold and results in the decrement of GP-region. For the La-substituted compounds, the $Mn^{3+}$ ions are substituted with the $Mn^{4+}$ ions (c.f. Fig. 5(d)). Thus, we can consider a transition from Fig. 5(g) to (f), where the replacement of $Mn^{3+}$ ion with $Mn^{4+}$ ion increases the cluster formation. However, as there are two structural phases in the La-substituted compounds, there is a possibility that the phase inhomogeneity gives rise to the observed GP behaviour.

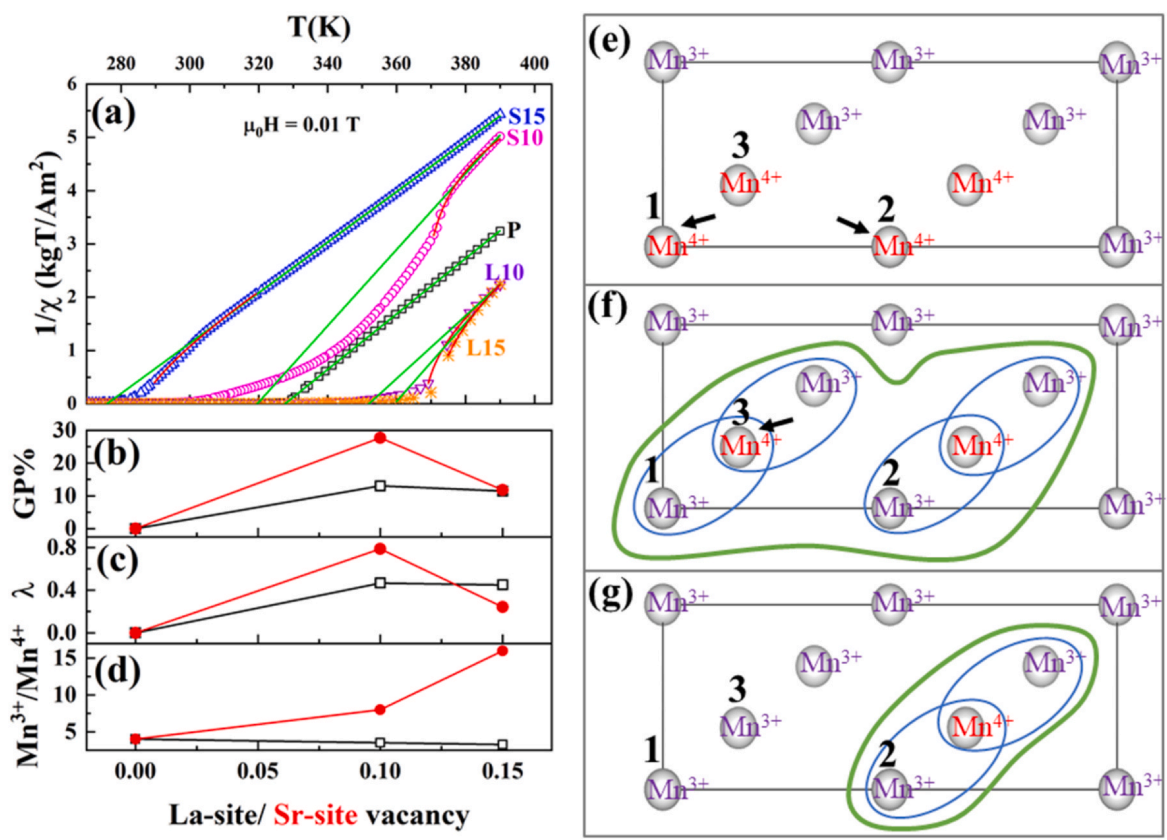

**Fig. 5.** (a) $1/\chi$ versus temperature. Calculated values of (b) GP%, (c) $\lambda$, and (d) ratio of JT active $Mn^{3+}$ to non-JT active $Mn^{4+}$ ions calculated using model M2C2; La-site (hollow black symbols) and Sr-site (solid red symbols) vacancy. (e)-(g) GP formation model.

## 4. Conclusion

The chemical, structural, electronic, magnetic and magnetocaloric properties of a series of $La_{0.8-x}Sr_{0.2-y}Mn_{1+x+y}O_3$ compounds have been studied here. The presence of excess Mn and chemical composition have been verified by the ion beam analysis. In the La-site substituted compound an additional orthorhombic phase is observed along with the principle rhombohedral phase. From the analysis of the Mn valence states and the structural phase formation, it has been confirmed that without external pressure (during synthesis) the smaller cations $Mn^{2+}/Mn^{3+}$ will not occupy the A-site of the perovskite, rather the $Mn_3O_4$-phase will form. The same analysis also confirms that the vacancy in the A-site will result in a similar vacancy in the B-site of the perovskite. La and Sr substitutions change $T_C$ in opposite ways with respect to the parent compound, this variation is explained in terms of the charge carrier bandwidth ($W$) in the compound. In the substituted compounds, a quenched disorder has been observed and has been characterized as GP disorder. The GP singularity did not show a simple linear relationship with the number JT-active ions, rather the results in this respect indicate a liner relationship of GP formation and the JT-active ions below a threshold limit at which the dimeron ($Mn^{3+}$-$Mn^{4+}$ pair) formation in the compound is maximum. Above this dimeron-threshold an inverse relationship of the GP formation and the JT-active ions is observed. A comparatively large value of the isothermal entropy change was observed for the Sr-substituted compounds while the La-substituted compounds exhibited large RCP values. Thus, this class of manganite perovskites with excess Mn, shows a unique way to tune the magnetic transition temperature and the isothermal entropy change near room temperature.

## Declaration of Competing Interest

The authors declare that they have no known competing financial interests or personal relationships that could have appeared to influence the work reported in this paper.

## Acknowledgments

The Swedish Foundation for Strategic Research (SSF, contract EM-16-0039) supporting research on materials for energy applications is gratefully acknowledged. Infrastructural grants by VR-RFI (#2017–00646_9) and SSF (contract RIF14-0053) supporting accelerator operation are gratefully acknowledged.

## Data Availability

The data that support the findings of this study are available upon reasonable request from the authors.

## Appendix A. Supporting information

Supplementary data associated with this article can be found in the online version at doi:10.1016/j.jallcom.2021.162714.

## References

[1] A.A. Belik, W. Yi, High-pressure synthesis, crystal chemistry and physics of perovskites with small cations at the A site, J. Phys. Condens. Matter 26 (2014) 163201.
[2] C.I. Thomas, M.R. Suchomel, G.V. Duong, A.M. Fogg, J.B. Claridge, M.J. Rosseinsky, Structure and magnetism of the A site scandium perovskite (Sc 0.94Mn0.06)

Mn0.65Ni0.35O 3 synthesized at high pressure, Philos. Trans. R. Soc. A Math. Phys. Eng. Sci. 372 (20130012) (2014).
[3] J. Cong, K. Zhai, Y. Chai, D. Shang, D.D. Khalyavin, R.D. Johnson, D.P. Kozlenko, S.E. Kichanov, A.M. Abakumov, A.A. Tsirlin, L. Dubrovinsky, X. Xu, Z. Sheng, S.V. Ovsyannikov, Y. Sun, Spin-induced multiferroicity in the binary perovskite manganite Mn2O3 *Nat*, Commun 9 (2018) 2996.
[4] A.V. Pashchenko, V.P. Pashchenko, Y.F. Revenko, V.K. Prokopenko, A.A. Shemyakov, G.G. Levchenko, N.E. Pismenova, V.V. Kitaev, Y.M. Gufan, A.G. Silcheva, V.P. Dyakonov, Structure, phase transitions, 55Mn NMR, 57Fe Mössbauer studies and magnetoresistive properties of La 0.6Sr 0.3Mn 1.1 - xFexO 3, J. Magn. Magn. Mater. 369 (2014) 122–126.
[5] A.V. Pashchenko, V.P. Pashchenko, N.A. Liedienov, V.K. Prokopenko, Y.F. Revenko, N.E. Pismenova, V.V. Burhovetskii, V.Y. Sycheva, A.V. Voznyak, G.G. Levchenko, V.P. Dyakonov, H. Szymczak, Structure, phase transitions, 55Mn NMR, magnetic and magnetotransport properties of the magnetoresistance La0.9-xAgxMn1.1O3-δ ceramics, J. Alloy. Compd. 709 (779–88) (2017).
[6] N.A. Liedienov, A.V. Pashchenko, V.P. Pashchenko, V.K. Prokopenko, Y.F. Revenko, A.S. Mazur, V.Y. Sycheva, V.I. Kamenev, G.G. Levchenko, Structure defects, phase transitions, magnetic resonance and magneto-transport properties of La0.6-xEuxSr0.3Mn1.1O3-δ ceramics, Low. Temp. Phys. 42 (1102–11) (2016).
[7] A.V. Pashchenko, V.P. Pashchenko, V.K. Prokopenko, Y.F. Revenko, N.G. Kisel, V.I. Kamenev, A.G. Sil'cheva, N.A. Ledenev, V.V. Burkhovetskii, G.G. Levchenko, Structural and magnetic inhomogeneities, phase transitions, 55Mn nuclear magnetic resonance, and magnetoresistive properties of La0.6 -xNdxSr0.3Mn1.1O3-δ ceramics, Phys. Solid State 56 (955–66) (2014).
[8] A.V. Pashchenko, V.P. Pashchenko, V.K. Prokopenko, Y.F. Revenko, A.S. Mazur, V.V. Burchovetskii, V.A. Turchenko, N.A. Liedienov, V.G. Pitsyuga, G.G. Levchenko, V.P. Dyakonov, H. Szymczak, The role of structural and magnetic in-homogeneities in the formation of magneto-transport properties of the La0.6-xSmxSr0.3Mn1.1O3-δ ceramics, J. Magn. Magn. Mater. 416 (457–65) (2016).
[9] A.V. Pashchenko, N.A. Liedienov, V.P. Pashchenko, V.K. Prokopenko, V.V. Burhovetskii, A.V. Voznyak, I.V. Fesych, D.D. Tatarchuk, Y.V. Didenko, A.I. Gudymenko, V.P. Kladko, A.A. Amirov, G.G. Levchenko, Modification of multifunctional properties of the magnetoresistive La0.6Sr0.15Bi0.15Mn1.1-xBxO3-δ ceramics when replacing manganese with 3d-ions of Cr, Fe, Co, Ni, J. Alloy. Compd. 767 (1117–25) (2018).
[10] A.A. Novokhatska, G.Y. Akimov, Role of excess manganese in formation of the structure and transport properties of manganite (Nd0.67Sr0.33)1-xMn1 + xO3 (x = 0, 0.2) sintered at 1273–1473 K, Phys. Solid State 60 (1394–7) (2018).
[11] E. Zubov, A. Pashchenko, N. Nedelko, I. Radelytskyi, K. Dyakonov, A. Krzyzewski, A. Slawska-Waniewska, V. Dyakonov, H. Szymczak, Magnetic and magnetoca-loric properties of the La0.9-xAgxMn1.1O3 compounds, Low Temp. Phys 43 (1190–5) (2017).
[12] Novokhatska A., Akimov G., Prylypko S., Revenko Y. and Burkhovetsky V. 2013 Evolution of microstructure and magnetoresistive properties of (La 0.65Sr0.35)0. 8Mn1.2O 3±Δ ceramics sintered at 800–1500°C *J. Appl. Phys.* **113** 2291.
[13] A.V. Pashchenko, V.P. Pashchenko, V.K. Prokopenko, Y.F. Revenko, Y.S. Prylipko, N.A. Ledenev, G.G. Levchenko, V.P. Dyakonov, H. Szymczak, Influence of structure defects on functional properties of magnetoresistance (Nd0.7Sr0.3)1-xMn1+xO3 ceramics, Acta Mater. 70 (218–27) (2014).
[14] V.P. Dyakonov, I. Fita, E. Zubov, V. Pashchenko, V. Mikhaylov, V. Prokopenko, Y. Bukhantsev, M. Arciszewska, W. Dobrowolski, A. Nabialek, H. Szymczak, Canted spin structure in clusters of the (La0.7Ca0.3)1-xMn1+xO3 perovskites, J. Magn. Magn. Mater. 246 (40–53) (2002).
[15] V. Dyakonov, A. Lawska-Waniewska, J. Kazmierczak, K. Piotrowski, O. Iesenchuk, H. Szymczak, E. Zubov, S. Myronova, V. Pashchenko, A. Pashchenko, A. Shemjakov, V. Varyukhin, S. Prilipko, V. Mikhaylov, Z. Kravchenko, A. Szytua, W. Bazela, Nanoparticle size effect on the magnetic and transport properties of (La0.7 Sr0.3)0.9 Mn1.1 O3 manganites, Low. Temp. Phys. 35 (568–76) (2009).
[16] Dul M. 2009 X-Ray Diffraction and Magnetic Properties Measurement of Nanopowder (La0.7Sr0.3)0.9Mn1.1O3 MANGANITE *Czas. Tech. Nauk. Pod.* 123–37.
[17] A.V. Pashchenko, V.P. Pashchenko, V.K. Prokopenko, A.G. Sil'cheva, Y.F. Revenko, A.A. Shemyakov, N.G. Kisel', V.P. Komarov, V.Y. Sycheva, S.V. Gorban', V.G. Pogrebnyak, Imperfection of the clustered perovskite structure, phase transitions, and magnetoresistive properties of ceramic La 0.6Sr 0.2Mn 1.2-xNi xO 3 ± δ (x = 0–0.3), Phys. Solid State 54 (2012) 767–777.
[18] É.E. Zubov, R. Puzhnyak, V.P. Pashchenko, V.I. Mikhaĭlov, A. Esenchuk, S.F. Mironova, S. Pekhota, V.P. Dyakonov, V.N. Varyukhin, H. Szymczak, Magnetocaloric effect in (La0.6Ca0.4)0.9Mn1.1O3, Phys. Solid State 51 (2090–4) (2009).
[19] R. Skini, S. Ghorai, P. Ström, S. Ivanov, D. Primetzhofer, P. Svedlindh, Large room temperature relative cooling power in La0.5Pr0.2Ca0.1Sr0.2MnO3, J. Alloy. Compd. 827 (2020) 154292.
[20] S. Ghorai, S.A. Ivanov, R. Skini, P. Ström, P. Svedlindh, Effect of reduced local lattice disorder on the magnetic properties of B-site substituted La0.8Sr0.2MnO3, J. Magn. Magn. Mater. 529 (167893) (2021).

[21] Z. Wei, N.A. Liedienov, Q. Li, A.V. Pashchenko, W. Xu, V.A. Turchenko, M. Yuan, I.V. Fesych, G.G. Levchenko, Influence of post-annealing, defect chemistry and high pressure on the magnetocaloric effect of non-stoichiometric La0.8-xK0.2Mn1+xO3 compounds, Ceram. Int. (2021).
[22] M. Gaudon, C. Laberty-Robert, F. Ansart, P. Stevens, A. Rousset, Preparation and characterization of La1-xSrxMnO3+δ (0 ≤ x ≤ 0.6) powder by sol-gel processing, Solid State Sci. 4 (2002) 125–133.
[23] A.A. Coelho, TOPAS and TOPAS-Academic: An optimization program integrating computer algebra and crystallographic objects written in C++, J. Appl. Crystallogr 51 (2018) 210–218.
[24] H.M. Rietveld, A profile refinement method for nuclear and magnetic structures, J. Appl. Crystallogr 2 (1969) 65–71.
[25] M. Mayer, SIMNRA, a simulation program for the analysis of NRA, RBS and ERDA, AIP Conference Proceedings vol 475, AIP, 2008, pp. 541–544.
[26] H.M. Rietveld, The Rietveld method ed International Union of Crystallograhy, *Phys. Scr.* 89 Oxford University Press, Oxford; New York, 2014.
[27] P. Radaelli, G. Iannone, M. Marezio, Structural effects on the magnetic and transport properties of perovskite 0.30, Phys. Rev. B - Condens. Matter Mater. 56 (1997) 8265–8276.
[28] P. Orgiani, A. Galdi, C. Aruta, V. Cataudella, G. De Filippis, C.A. Perroni, V. Marigliano Ramaglia, R. Ciancio, N.B. Brookes, M. Moretti Sala, G. Ghiringhelli, L. Maritato, Multiple double-exchange mechanism by Mn$^{2+}$ doping in manganite compounds *Phys. Rev. B - Condens. Matter Mater*, Phys. Rev. B Condens. Matter. 82 (2010).
[29] E. Beyreuther, S. Grafström, L.M. Eng, C. Thiele, K. Dörr, XPS investigation of Mn valence in lanthanum manganite thin films under variation of oxygen content, Phys. Rev. B 73 (2006) 155425.
[30] S. Ghorai, S.A. Ivanov, R. Skini, P. Svedlindh, Evolution of Griffiths phase and critical behaviour of La 1-x Pb x MnO 3 ± y solid solutions, J. Phys. Condens. Matter 33 (2021) 145801.
[31] P.G. Radaelli, G. Iannone, M. Marezio, H.Y. Hwang, S.-W. Cheong, J.D. Jorgensen, D.N. Argyriou, Structural effects on the magnetic and transport properties of perovskite A1-xA'xMnO3 (x=0.25, 0.30), Phys. Rev. B 56 (1997) 8265–8276.
[32] K. Li, R. Cheng, S. Wang, Y. Zhang, Infrared transmittance spectra of the granular perovskite La2/3Ca1/3MnO3, J. Phys. Condens. Matter 10 (1998) 4315–4322.
[33] M.H. Phan, S.C. Yu, Review of the magnetocaloric effect in manganite materials, J. Magn. Magn. Mater. 308 (2007) 325–340.
[34] W. Chen, W. Zhong, D.L. Hou, R.W. Gao, W.C. Feng, M.G. Zhu, Y.W. Du, Preparation and magnetocaloric effect of self-doped La0.8-xNa0.2€xMnO3+δ (€ = vacancies) polycrystal, J. Phys. Condens. Matter 14 (11889–96) (2002).
[35] J. Lyubina, Magnetocaloric materials for energy efficient cooling, J. Phys. D. Appl. Phys. 50 (2017) 53002.
[36] T.D. Shen, R.B. Schwarz, J.Y. Coulter, J.D. Thompson, Magnetocaloric effect in bulk amorphous Pd 40Ni 22.5Fe 17.5P 20 alloy, J. Appl. Phys. 91 (2002) 5240–5245.
[37] J.Y. Law, V. Franco, L.M. Moreno-Ramírez, A. Conde, D.Y. Karpenkov, I. Radulov, K.P. Skokov, O. Gutfleisch, A quantitative criterion for determining the order of magnetic phase transitions using the magnetocaloric effect, Nat. Commun. 9 (2018) 2680.
[38] R. Skini, M. Khlifi, E.K. Hlil, An efficient composite magnetocaloric material with a tunable temperature transition in K-deficient manganites, RSC Adv. 6 (2016) 34271–34279.
[39] S. Bouzidi, M.A. Gdaiem, J. Dhahri, E.K. Hlil, Large magnetocaloric entropy change at room temperature in soft ferromagnetic manganites, RSC Adv. 9 (2019) 65–76.
[40] S. Ghorai, R. Skini, D. Hedlund, P. Ström, P. Svedlindh, Field induced crossover in critical behaviour and direct measurement of the magnetocaloric properties of La0.4Pr0.3Ca0.1Sr0.2MnO3, Sci. Rep. 10 (2020) 19485.
[41] R. Skini, A. Omri, M. Khlifi, E. Dhahri, E.K. Hlil, Large magnetocaloric effect in lanthanum-deficiency manganites La0.8-x☐xCa0.2MnO3 (0.00≤x≤0.20) with a first-order magnetic phase transition, J. Magn. Magn. Mater. 364 (5–10) (2014).
[42] M. Wali, R. Skini, M. Khlifi, E. Dhahri, E.K. Hlil, A giant magnetocaloric effect with a tunable temperature transition close to room temperature in Na-deficient La 0.8 Na 0.2−x ☐ x MnO 3 manganites *Dalt*, Trans 44 (2015) 12796–12803.
[43] V.K. Pecharsky, K.A. Gschneidner, Giant magnetic effect in Gd5(Si2Ge2), Phys. Rev. Lett. 78 (1997) 4494–4497.
[44] C. Magen, P.A. Algarabel, L. Morellon, J.P. Araãjo, C. Ritter, M.R. Ibarra, A.M. Pereira, J.B. Sousa, Observation of a griffiths-like phase in the magnetoca-loric compound Tb5Si2Ge2, Phys. Rev. Lett. 96 (2006).
[45] A.K. Pramanik, A. Banerjee, Griffiths phase and its evolution with Mn-site dis-order in the half-doped manganite Pr 0.5 Sr 0.5 Mn 1-y Ga y O 3 ( y=0.0, 0.025, and 0.05), Phys. Rev. B Condens. Matter Mater. 81 (2010) 024431.
[46] L. Downward, F. Bridges, S. Bushart, J.J. Neumeier, N. Dilley, L. Zhou, Universal relationship between magnetization and changes in the local structure of La1-xCaxMnO3: evidence for magnetic dimers, Phys. Rev. Lett. 95 (2005).